\documentclass[aps,prb,twocolumn,superscriptaddress,showpacs,floatfix]{revtex4}
\usepackage{graphicx}
\usepackage{amssymb,amsmath}
\usepackage{longtable}

\newcommand{\bQ}{\mbox{\boldmath$Q$}}

\begin{document}


\title{Absence of localized-spin magnetism in the narrow-gap semiconductor FeSb$_{2}$.}


\author{I.~A.~Zaliznyak}
\email[Email to: ]{zaliznyak@bnl.gov}
\affiliation{CMPMSD,
Brookhaven National Laboratory, Upton, New York 11973-5000}

\author{A.~T.~Savici}
\affiliation{Department of Physics and Astronomy, The Johns Hopkins University, Baltimore, Maryland 21218}
\affiliation{
Neutron Scattering Science Division,
Oak Ridge National Laboratory, Oak Ridge, Tennessee 37831, USA}

\author{V. O.~Garlea}
\affiliation{
Neutron Scattering Science Division,
Oak Ridge National Laboratory, Oak Ridge, Tennessee 37831, USA}

\author{Rongwei~Hu}
\altaffiliation[Present address: ]{Ames Laboratory, US DOE, and Department of Physics and Astronomy, Iowa State University, Ames, IA 50011}
\affiliation{CMPMSD,
Brookhaven National Laboratory, Upton, New York 11973-5000}

\author{C.~Petrovic}
\affiliation{CMPMSD,
Brookhaven National Laboratory, Upton, New York 11973-5000}


\date{\today}

\begin{abstract}
We report the inelastic neutron scattering measurements aimed at investigating the origin of temperature-induced paramagnetism in narrow-gap semiconductor FeSb$_2$. We find that inelastic response for energies up to 60 meV and at temperatures $\approx 4.2$ K, $\approx 300$ K and $\approx 550$ K is essentially consistent with the scattering by lattice phonon excitations. We observe no evidence for a well-defined magnetic peak corresponding to the excitation from the non-magnetic S = 0 singlet ground state to a state of magnetic multiplet in the localized spin picture. Our data establish the quantitative limit of $S_{eff}^2 \lesssim 0.25$ on the fluctuating local spin. However, a broad magnetic scattering continuum in the 15 meV to 35 meV energy range is not ruled out by our data. Our findings make description in terms of the localized Fe spins unlikely and suggest that paramagnetic susceptibility of itinerant electrons is at the origin of the temperature-induced magnetism in FeSb$_2$.

\end{abstract}


\maketitle


\section{Introduction}

A tight balance between strong covalent hybridization, the tendency to band delocalization, strong electronic correlation and the crystal field potential leads to a rich variety of unusual electronic states in the inter-metallic pnictides and chalcogenides TM$_x$, where T is the transition metal (Mn, Fe, Co, Ni, ...), M is a metalloid (Sb, As, Te, ...), and $0.5\lesssim x \lesssim 3$.
\cite{Neel_1952,Goodenough_JSolStChem1972,Holseth_etal_Acta1970,Kjekshus_Acta1979,Yamaguchi_JMMM1992,RadhakrishnaCable_PRB1996,
YashiroYamaguchi_JPSJ1973,Yamaguchi_JPSJ1978,YamaguchiWatanabe_JPSJ1978,DeGroot_PRL1983,Mavropoulos_JPCM2007,
WadaTanabe_APL2001,Ohno_PRB2007,Benthien_EPL2007,Sun_APE2009,Hu_PRB2008,Hu_APL2008,Hu_PRB2009,
Chen_etal_Nature2008,Izyumov_Uspekhi2008,Christianson_Nature2008,Lumsden_PRL2009,Christianson_PRL2009,Diallo_PRL2009,Zhao_NatPhys2009}%
Among them are semi-metallic and semiconducting compounds with rich variety of magnetic properties and magnetic structures, \cite{Neel_1952,Goodenough_JSolStChem1972,Holseth_etal_Acta1970,Kjekshus_Acta1979,Yamaguchi_JMMM1992,RadhakrishnaCable_PRB1996,
YashiroYamaguchi_JPSJ1973,Yamaguchi_JPSJ1978,YamaguchiWatanabe_JPSJ1978} half-metallic ferromagnets, \cite{DeGroot_PRL1983,Mavropoulos_JPCM2007} materials showing giant magnetocaloric effect, \cite{WadaTanabe_APL2001} unusually high thermopower, \cite{Ohno_PRB2007,Benthien_EPL2007,Sun_APE2009} colossal magneto-resistance and giant carrier mobility. \cite{Hu_PRB2008,Hu_APL2008,Hu_PRB2009}
Many of the unique electronic properties which are of great current and future technological interest are intertwined with the emerging magnetism in these materials. Recent discovery of the new variety of the high-temperature superconductivity in layered transition metal pnictides, which is also closely related with magnetism, \cite{Chen_etal_Nature2008,Izyumov_Uspekhi2008,Christianson_Nature2008,Lumsden_PRL2009,Christianson_PRL2009,Diallo_PRL2009,Zhao_NatPhys2009} has generated new surge of interest in the physical properties of these compounds.

In defiance of the na\"{i}ve expectation that alloys composed of metallic constituents should be metallic, many pnictides and chalcogenides are actually semiconducting, or even insulating. Non-metallic and half-metallic behaviors result primarily from the hybridization gaps imposed by strong covalent bonding between the transition metal $3d$ and the metalloid $p$ (or, more precisely, $sp$) orbitals, which dominate the electronic band structure near the Fermi level. These are also further enhanced by strong correlation of the narrow-band $d$-electrons. \cite{Goodenough_JSolStChem1972,WeinertWatson_PRB1998,Lukoyanov_EPJ2006,KunesAnisimov_PRB2008}

Iron diantimonide FeSb$_2$ is a narrow gap semiconductor representative of the TM$_2$ pnictide family with a number of interesting and unusual behaviors. \cite{Ohno_PRB2007,Benthien_EPL2007,Sun_APE2009,Hu_PRB2008,Hu_APL2008,Hu_PRB2009,Petrovic_PRB2003,Hu_PRB2007,Petrovic_PRB2005} Perhaps most outstanding are the temperature-induced paramagnetism similar to that found in iron silicide, FeSi, \cite{Tajima_PRB1988} and strongly anisotropic electrical transport properties. \cite{Petrovic_PRB2003,Hu_PRB2007,Petrovic_PRB2005} Below room temperature and down to about 40 K the conductivity of single crystal FeSb$_2$ shows metallic behavior along one of the crystallographic directions (the resistivity for current along the $c$-axis decreases with the decreasing temperature) and semiconducting activated behaviors along two other, $a$ and $b$ axes \cite{Erratum}. Such behavior suggests anisotropic band structure, which may results from a particular network of hybridized orbitals, a flavor of the orbital order. Then, either there is a one-dimensional metal-insulator transition changing the $c$-axis transport around 40 K, or a sequence of semiconductor cross-overs corresponding to anisotropic band gaps, which vary from about 4 meV for transport along the $c$ axis to about 30 meV perpendicular to it.

As conductivity of FeSb$_2$ increases with the increasing temperature, an excess Shottky-like electronic heat capacity \cite{Westrum_RCR1979} and an exponentially-activated paramagnetic susceptibility \cite{Petrovic_PRB2003,Hu_PRB2007,Petrovic_PRB2005,Fan_JSSC1972} also develop. The origin of these and other intriguing properties, such as the giant thermopower, colossal magneto-resistance and high carrier mobilities, lies in the physical nature of the lowest unoccupied and the highest occupied electronic states near the Fermi level. Currently, the structure and the composition of these states are unclear. In particular, the temperature-induced paramagnetic susceptibility could be analyzed either in the framework of the ionic, localized-spin model, or as a Pauli paramagnetism of itinerant electrons belonging to two bands separated by a narrow gap containing the Fermi energy, $E_F$. \cite{Petrovic_PRB2005} In the former, localized-spin picture, it involves thermally activated spin state transition between the non-magnetic S = 0 and magnetic S=1 states of the crystal-field split multiplet of Fe $3d$ electronic levels, corresponding to semi-localized narrow bands. \cite{Petrovic_PRB2003} In this scenario, the corresponding crystal-field spin excitation should be observable in the inelastic magnetic neutron scattering experiment. Search for this excitation aimed at clarifying the nature (localized vs itinerant) of the temperature-induced magnetism and thus the lowest unoccupied electronic states in FeSb$_2$, was the goal of the present study.

Fitting the temperature-dependent magnetic susceptibility $\chi(T)$ to a thermally induced spin-state transition in the localized spin model yields a spin gap $\Delta_\chi \approx 47$ meV and an effective magnetic moment of about $1.2 \mu_B$ (Bohr magnetons) for the magnetic high-spin state. \cite{Petrovic_PRB2003} The corresponding singlet-triplet spin excitation would therefore be expected to produce a peak of significant intensity in the FeSb$_2$ magnetic neutron scattering cross-section at an energy $\approx \Delta_\chi$. The intensity of such an excitation is expected to show characteristic temperature dependence. Namely, it should decrease with the increasing temperature, as the transition probability gets depleted with the increasing thermal population of excited magnetic states. \cite{Sasago_PRB1997} We have conducted an extensive search for such magnetic spin excitation for energies up to 60 meV and at temperatures 10 K, 300 K and 550 K using inelastic thermal neutron scattering, which is reported here. 

The paper is organized as follows. Section 2 contains description of the FeSb$_2$ crystal structure and discussion of its relation with the electronic properties of this material (the detailed qualitative discussion of band hybridization aspects illustrated in Fig. \ref{Fig0-structure} follows Goodenough's paper [\onlinecite{Goodenough_JSolStChem1972}]; it is rather specialized and could be skipped on the first reading). Experimental details are described in Section 3, the resulting data and data analysis are presented in Section 4, which is followed by the brief summary of our conclusions in Section 5. Appendix A presents details of the quantitative comparison of magnetic and phonon neutron scattering intensities.

\section{F\lowercase{e}S\lowercase{b}$_2$ crystal and electronic structure}

\begin{figure}[!t]
\begin{center}
\includegraphics[width=0.8\linewidth,angle=0]{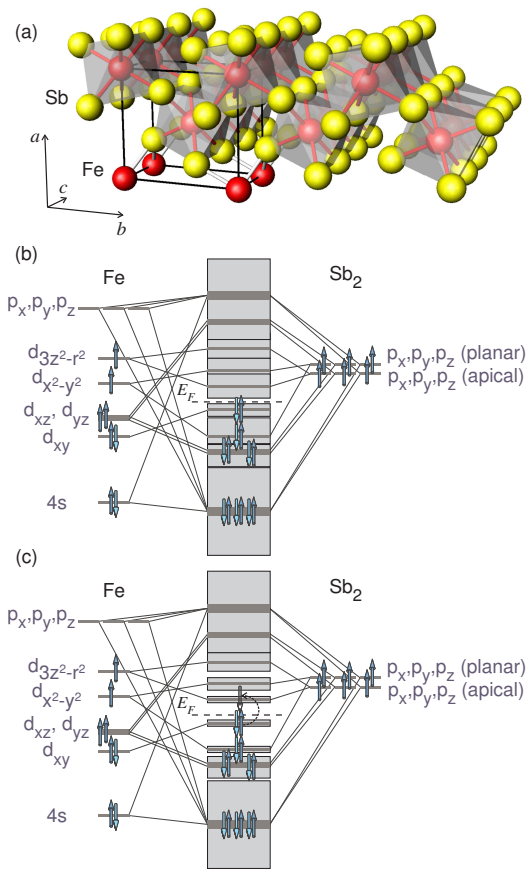}
\caption{\label{Fig0-structure} (a) Crystal structure of FeSb$_2$. Atoms beyond the unit cell are shown to illustrate the main structural motif -- Fe-Sb octahedra sharing edges along $c$ direction and corners along $a$ and $b$. (b), (c) Schematics of the electronic band structure of FeSb$_2$ resulting from the crystal field splitting and strong covalent hybridization. (c) illustrates the spin-state transition and the corresponding crystal field excitation in the quasi-localized narrow-band scenario.}
\end{center}
\vspace{-0.2in}
\end{figure}

Iron diantimonide FeSb$_2$ crystalizes in a FeS$_2$ marcasite structure shown in Figure \ref{Fig0-structure} (a). \cite{Kjekshus_Acta1979,Petrovic_PRB2003,Petrovic_PRB2005} It features chains of edge-sharing FeSb$_6$ octahedra running along the $c$ axis, which are stacked in a somewhat corrugated body-centered geometry, sharing corners in $a$ and $b$ directions. The lattice is orthorhombic (space group $Pnnm$), with two formula units per unit cell and room temperature lattice parameters $a \approx 5.83$ \AA, $b \approx 6.53$ \AA\ and $c \approx 3.2$ \AA.

%
%

The electronic structure of Fe $3d$ levels in FeSb$_2$ is determined by the hybridization with the Sb $5p$ and $5s$ orbitals (or $sp^3$ hybrid orbitals, but we retain $p_x, p_y, p_z$ tags for simplicity) and by the crystal field of the Sb octahedral environment. At low temperature, Sb octahedra are slightly squashed, with two shorter (``apical'', $\approx 2.56$ \AA\ below $\approx 100$ K) and four longer (``planar'', $\approx 2.59$ \AA\ below $\approx 100$ K) Fe-Sb bonds. \cite{Petrovic_PRB2005} Such distortion is opposite to that found in the family of high-temperature superconducting layered perovskite cupric oxides and their nickel- or cobalt-based relatives, where oxygen octahedra are elongated and tetragonal crystal field favors $d_{yz,zx}$ and $d_{3z^2-r^2}$ orbitals. Schematics of the resulting band structure following from qualitative arguments of Ref. \onlinecite{Goodenough_JSolStChem1972} is illustrated in Fig. \ref{Fig0-structure} (b), (c).

In discussing the local Fe 3$d$ electronic structure, it is convenient to associate the long and the short bond directions with (X, Y) and Z coordinate axes, respectively. In this notation, the tetragonal crystal field of squashed octahedra splits the $e_g$ and $t_{2g}$ multiplets so that $d_{x^2-y^2}$ and $d_{xy}$ orbitals have lower energy compared to $d_{3z^2-r^2}$ and $d_{yz,zx}$ orbitals, respectively. Such level hierarchy is opposite to that used in Ref. \onlinecite{Goodenough_JSolStChem1972} for constructing schematic band structure of marcasite iron pnictides starting from the ionic Fe$^{4+}$ $3d^4$ configuration. This distinction, however, is unimportant, since distortion of the octahedra is quite small and further decreases upon heating, essentially vanishing at T $\approx 500$ K. \cite{Petrovic_PRB2005} Hence, the electronic band structure is mainly determined by covalent shifts resulting from $3d - 5p$ hybridization and formation of bonding and antibonding states. This pushes $d_{xz, yz}$-derived states down, perhaps below the non-bonding or weakly-bonding $d_{xy}$ level, so that splitting of the $t_{2g}$ multiplet is the same as for the crystal field of elongated octahedra.

An interesting problem arises with assigning an ionic state, such as $3d^4$, to Fe. With such an assumption, the Fermi level lies within the $t_{2g}$ multiplet and the lowest unoccupied states arise from the essentially non-bonding $d_{xy}$ orbital forming a narrow, semi-localized conduction band. \cite{Goodenough_JSolStChem1972}
Such scenario could in fact explain the anisotropic conductance experimentally observed in FeSb$_2$. \cite{Petrovic_PRB2003} Indeed, a band derived from weakly overlapping nearly non-bonding $d_{xy}$ orbitals could give rise to nearly one-dimensional conductance along the $c$-axis direction, Fig. \ref{Fig0-structure}.
Such assignment, however, disagrees with the recent $LDA+U$ band structure calculations, \cite{KunesAnisimov_PRB2008,Lukoyanov_EPJ2006} which find that the Fermi level lies above $t_{2g}$ states, and the conduction band is mainly of $d_{3z^2-r^2}$ origin, as shown in Fig. \ref{Fig0-structure} (b), (c). Ionic picture also disagrees with the general expectation of strong $3d-5p$ covalent Fe-Sb bonding in a metal-metalloid compound. It is also possible that covalent shifts are even stronger than those shown in Fig. \ref{Fig0-structure} (b), (c), so that $d_{3z^2-r^2}$ and $d_{x^2-y^2}$ levels appear below the $d_{xy}$ level. While the precise level hierarchy is presently unclear, it does not impact the arguments distinguishing between the itinerant and localized-spin magnetism presented below.

In the hybridization-dominated covalent band scheme of FeSb$_2$ illustrated in Fig. \ref{Fig0-structure} (b), (c), the $d_{x^2-y^2}$-derived valence and $d_{3z^2-r^2}$-derived conduction bands  agree with the $LDA+U$ results of Refs. \onlinecite{Lukoyanov_EPJ2006}, \onlinecite{KunesAnisimov_PRB2008}. The difference between panels (b) and (c) is in the bandwidth of the valence and the conduction bands near the Fermi energy, $E_F$. In the case of Fig. \ref{Fig0-structure} (b), thermally excited electrons and holes are strongly itinerant and exhibit a weak Pauli paramagnetism, similar to the picture proposed for FeSi.\cite{Petrovic_PRB2005,Jaccarino_PhysRev1967} There is no local spins and local magnetic moments in this scenario, and only a weak and broad signal is expected in a magnetic neutron scattering experiment. This is consistent with the weak temperature-induced quasi-elastic paramagnetic scattering observed by neutrons in FeSi. \cite{Tajima_PRB1988} In the narrow-band scenario of panel (c) on the other hand, the strong on-site correlation dominates the electronic properties, electrons near the Fermi level are essentially localized, and temperature-induced magnetic state corresponds to a local S=1 triplet.
In the band picture language this can be viewed as a localized S = 1 electron-hole triplet bound state. A well-defined peak in magnetic neutron scattering at an energy corresponding to the local singlet-triplet spin excitation, and with the characteristic decrease in intensity with the increasing temperature, could then be expected. \cite{Sasago_PRB1997}

\section{Experimental procedure}

\begin{figure}[!h]
\begin{center}
\includegraphics[width=1.\linewidth,angle=0]{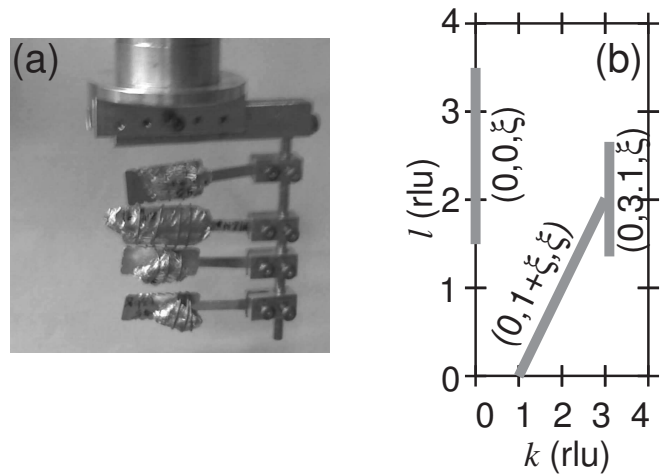}
\caption{\label{Fig1-sample} (a) All-aluminum sample holder with 4-crystal FeSb$_2$ assembly used in our measurements. (b) Scaled to \AA$^{-1}$ schematics of the $(0kl)$ reciprocal lattice zone. Grey bars show scan directions for maps shown in Figures \ref{Fig3-Sqw_maps} and \ref{Fig4-Imchi_qw_maps}.}
\end{center}
\vspace{-0.2in}
\end{figure}

Our FeSb$_2$ crystals were similar to those used in the previous studies reported in Refs. \onlinecite{Hu_APL2008}, \onlinecite{Petrovic_PRB2003}, \onlinecite{Petrovic_PRB2005}, and were grown from the excess antimony flux as described in Ref. \onlinecite{Petrovic_PRB2003}. Our sample was an array of four large single crystals with total mass $m = 7.46$ g mounted on an aluminum alloy (Al6061) sampleholder shown in Fig. \ref{Fig1-sample} (a). Crystals were co-aligned to within $0.7^\circ$ in the horizontal scattering plane, with $a$ lattice direction being vertical. The sample assembly shown in Fig. \ref{Fig1-sample} (a) was mounted in the closed-cycle refrigerator capable of maintaining temperature in the range from about 4 K to about 600 K. In all measurements the $(0,k,l)$ reciprocal lattice plane of the sample was kept in the horizontal scattering plane.

%
%

Neutron scattering measurements were performed using the HB1 thermal neutron triple axis spectrometer at the HFIR Center for Neutron Scattering at the Oak Ridge National Laboratory (ORNL). Monochromatic incident neutrons were obtained using the (002) reflection from vertically focussing pyrolytic graphite (PG) crystals, scattered neutrons were analyzed using similar PG(002) analyzer crystals. Two $1''$ thick PG filters were placed after the sample to suppress scattered beam contamination by higher order reflections in PG monochromator and analyzer. Neutron beam collimations were $\approx 48'-40'-60'-120'$, from reactor to detector. Two final scattered neutron energies were used, $E_f = 14.7$ meV for the high-resolution mode, and $E_f = 30.5$ meV for the low-resolution and high-intensity mode. In the latter case, the total volume of the sampled phase space increases by a factor between 4 and 5 in our energy and momentum transfer range. This results in a proportionally higher sensitivity to weakly dispersive features in the scattering cross-section. Broad surveys of scattering intensity for energy transfers up to 60 meV and at temperatures  $\approx 4.2$ K, $\approx 300$ K and $\approx 550$ K were performed using high-intensity configuration with $E_f = 30.5$ meV. Lines in Fig. \ref{Fig1-sample} (b) show the corresponding constant-energy scans in the $b^* - c^*$ reciprocal lattice plane.

%
\begin{figure}[b]
\begin{center}
\vspace{-0.5in}
\includegraphics[width=1.\linewidth,angle=0]{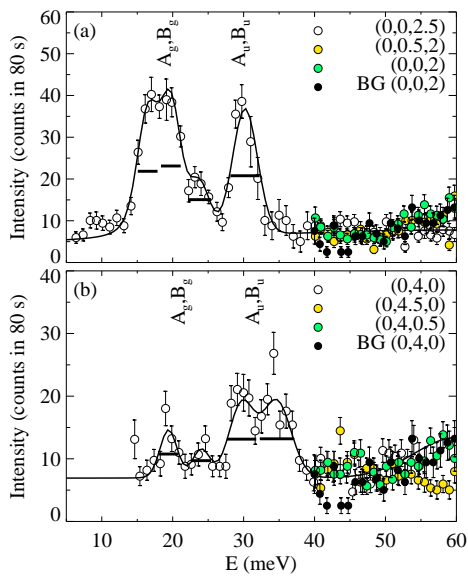}%
\vspace{-0.5in}
\caption{\label{Fig2-Escans} Constant-{\bQ} scans near \bQ\ =  (0, 0, 2), $\approx$ along the $c$-axis direction (a) and near \bQ\ = (0, 4, 0), $\approx$ along the $b$-axis (b), measured with $E_f = 14.7$ meV at T = 4.1(1) K. Black circles show background measured without the sample. Lines show fits to a number of resolution-corrected Lorentzian peaks describing phonon scattering. Horizontal bars show the calculated full width at half maximum (FWHM) of the instrument resolution, \cite{Cooper-Nathans} projected on the energy axis. }
\end{center}
\vspace{-0.2in}
\end{figure}
%

%
\begin{table*}[!t]
\vspace{-0.25in}
\caption{Intensity, position and the full width at half maximum (FWHM) of the resolution-corrected Lorentzian fits to four distinct phonon groups shown in Fig. \ref{Fig2-Escans}. FWHM $\lesssim 0.1$ meV is less than $10\%$ of the instrument energy resolution and implies no experimentally observable width.}%
\label{Table1}
 \vspace{0.05in}
\begin{tabular}[c]{lcccccccccccc}
 \hline\hline
Wave vector, & \multicolumn{4}{c}{Integral Intensity (cts$\cdot$meV)} &  \multicolumn{4}{c}{Position (meV)} &\multicolumn{4}{c}{FWHM (meV)} \\
 \cline{2-13}
\bQ (r. l. u.) & $I_1$ & $I_2$ & $I_3$ & $I_4$  & $E_1$ & $E_2$ & $E_3$ & $E_4$ & $W_1$ &  $W_2$ &  $W_3$ &  $W_4$ \\
 \hline
(0, 0, 2.5) & 133(10) & 130(10) & 41(6) & 134(12) & 16.5(2) & 19.7(2) & 23.8(3) & 30.2(2) & 1.5(4) & 1.2(4) & $\lesssim 0.1$ & $\lesssim 0.1$ \\
(0, 4, 0) & 24(5) & 14(5) & 53(7) & 63(7) & 19.4(3) & 24.0(5) & 29.7(3) & 34.7(3) & $\lesssim 0.1$ & $\lesssim 0.1$ &  $\lesssim 0.1$ & $\lesssim 0.1$ \\
 \hline\hline
\end{tabular}
\end{table*}
%

\section{Results and discussion}

We first searched for the well-defined crystal-field singlet-triplet excitation, which would be expected in the narrow-band localized-spin picture, by carrying out energy scans at different wave vector transfers in the high-resolution configuration with $E_f = 14.7$ meV. Several such scans for wave vectors \bQ\ near $b^*$ and $c^*$ directions are shown in Figure \ref{Fig2-Escans}. We observed no features which could be identified with the expected magnetic excitation in the 35 meV to 60 meV range. A slight increase in the background (BG) towards higher energies, which is more pronounced for smaller $Q$, is an instrumental effect associated with the detector vessel approaching the incident neutron beam. This was confirmed by background measurements with sample removed from the beam, which are shown by solid black circles in Figure \ref{Fig2-Escans} (a, b).

A number of peaks seen in the energy range from 10 to 35 meV are attributable to optic phonon modes. A plethora of optic phonons were observed at these energies by Raman and far-infrared (FIR) optical spectroscopy. \cite{Perucci_EPJ2006,Racu_JAP2008,Lazarevic_PRB2009} Six phonon modes at 18.7 meV, 19.0 meV ($A_g$ symmetry), and 11.2 meV ($B_{2g}$), 18.8 meV ($B_{3g}$), 19.1 meV and 21.6 meV ($B_{1g}$) were identified in Raman experiments. \cite{Racu_JAP2008,Lazarevic_PRB2009} Additional peaks, including high-energy modes, were observed in the FIR reflectivity measurements at 13.2 meV, 28.6 meV, 31.9 meV, 33.6 meV (electric field of light $\parallel b$) and 15.0 meV, 26.8 meV, 32.4 meV (electric field of light $\perp b$). Solid lines in Figure \ref{Fig2-Escans} (a, b) show resolution-corrected \cite{Cooper-Nathans} fits of our data to a number of Lorentzian peaks, whose parameters are listed in Table \ref{Table1}. Within the instrumental resolution shown by horizontal bars in Figure \ref{Fig2-Escans}, peak positions observed in our neutron scattering measurements agree well with Raman/FIR data. Significant intrinsic width of $E_1$ and $E_2$ peaks at $\bQ = (0,0,2.5)$ reflects the fact that there are several distinct phonon modes within each peak.

The phonon origin of peaks in Figure \ref{Fig2-Escans} is further supported by the temperature dependence of their intensity, which increases upon heating. Such behavior is indeed typical of phonons, whose contribution to scattering cross-section at a wave vector \bQ\ is given by the dynamical correlation function $S_p(\bQ, E)$ of an oscillator. It is related to the imaginary part of the oscillator dynamical susceptibility $\chi_p''(\bQ, E)$ through the fluctuation-dissipation theorem, \cite{CallenWelton_PhysRev1951,ZaliznyakLee_2005}
\begin{equation}
\label{FD_theorem}
\pi S_p(\bQ, E) = \chi_p''(\bQ, E)\left/ \left(1 - e^{-E/k_B T}\right)\right. .
\end{equation}
In the temperature range where $\chi_p''(\bQ, E)$ of the phonon oscillator does not change much (damping, anharmonicities, \emph{etc}, are small), the measured $S_p(\bQ, E)$ increases. This increase is due to the decrease of the detailed balance factor (often called Bose factor, since  $(1 - e^{-E/k_B T})^{-1} = 1 + n_B(T)$) in the denominator of Eq. \ref{FD_theorem}, and is most pronounced for $E \lesssim T$. The imaginary part of dynamical susceptibility of a damped oscillator can be represented as a difference of two Lorentzian peaks centered at $\pm E_0$ ($E_0/\hbar$ is the undamped oscillator frequency; this naturally satisfies the causality requirement, $\chi_p''(\bQ, E) = -\chi_p''(\bQ, -E)$). In cases like ours, where peak energy is much larger than damping, the contribution of the negative-energy peak can be neglected at $E \gtrsim T$, which justifies using simple Lorentzian line shapes.

%
\begin{figure}[!t]
\begin{center}
\includegraphics[width=3.2in,angle=0]{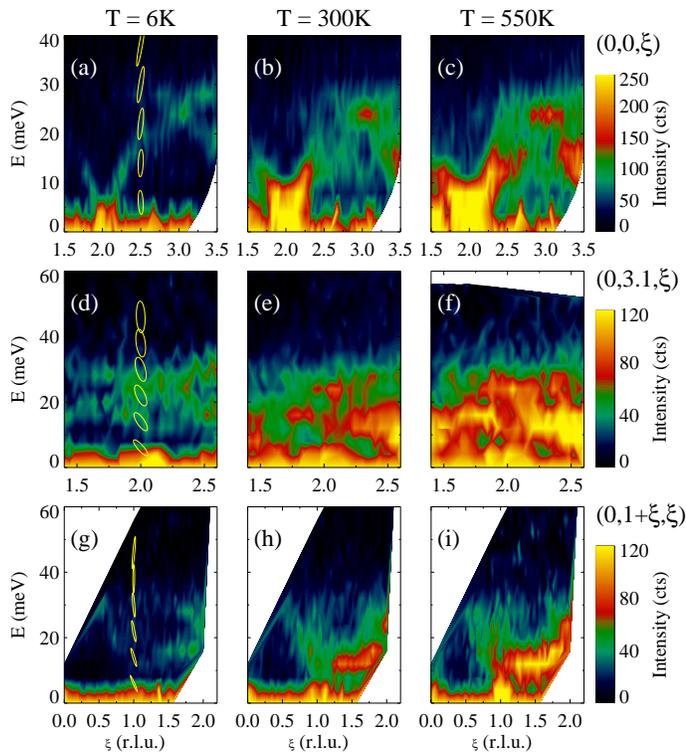}%
\caption{\label{Fig3-Sqw_maps} Contour map of neutron inelastic scattering intensity for $\bQ = (0, 0, \xi)$, (a-c), $\bQ = (0, 3.1, \xi)$, (d-f), and $\bQ = (0, 1 + \xi, \xi)$, (g-i), at T = 5 K, 300 K and 550 K (from left to right). Intensity is shown in counts per monitor count corresponding to counting time of $\approx 1.5$ minutes at 4 meV and $\approx 4$ minutes at 60 meV. Ellipses show the calculated full width at half maximum (FWHM) instrument resolution. \cite{Cooper-Nathans} }
\end{center}
\end{figure}
%

%
 \begin{figure}[!t]
 \includegraphics[width=3.2in,angle=0]{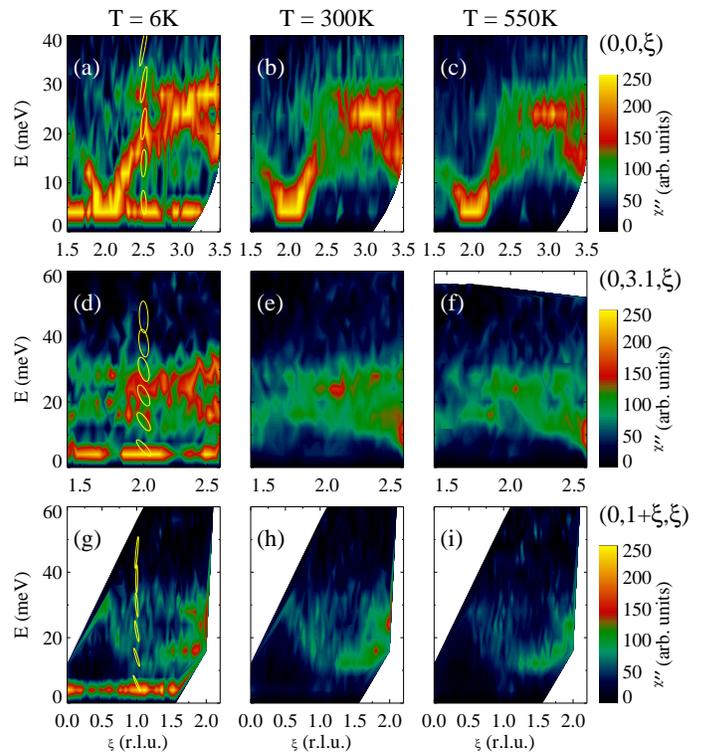}
 \caption{\label{Fig4-Imchi_qw_maps} Imaginary part of the dynamical susceptibility obtained by applying the detailed balance factor of Eq. \ref{FD_theorem} to the data shown in Figure \ref{Fig3-Sqw_maps}.}
 \end{figure}
%

Having established the phonon nature of dominant scattering features observed in FeSb$_2$ at low T in scans with $E_f = 14.7$ meV, we decided to perform broad surveys of scattering using the high-intensity mode with $E_f = 30.5$ meV. By studying its evolution with temperature we could then attempt to single out some evidence of the non-phonon magnetic scattering. To this end, we have measured scattering along $(0, 0, \xi)$, $(0, 3.1, \xi)$, and  $(0, 1+\xi, \xi)$ directions shown in Fig. \ref{Fig1-sample} (b), for energies up to 40 meV, or 60 meV. Corresponding color contour maps of the measured inelastic scattering intensity are shown in the three rows, (a-c), (d-f) and (g-i), of  Figure \ref{Fig3-Sqw_maps}. The data was collected in a set of constant-energy scans with step $dQ \approx 0.075$ \AA, taken every 2 meV. Three columns in the figure show intensities measured at three different temperatures, T = 6 K, 300 K and 550 K. Ellipses in panels (a), (d) and (g) illustrate the calculated full width at half maximum (FWHM) of the instrument resolution function projected along the energy axis. \cite{Cooper-Nathans}

Similar to the $E_f = 14.7$ meV data of Fig. \ref{Fig2-Escans}, the scattering in Fig. \ref{Fig3-Sqw_maps} is also dominated by phonons and increases strongly with the increasing temperature. There is no evidence for any scattering above the BG level in the 35 meV to 60 meV range. Both optic phonons and an acoustic phonon emerging from the $(0,0,2)$ Bragg peak are seen in  $(0, 0, \xi)$ maps of Fig. \ref{Fig3-Sqw_maps} (a-c), while for $(0, 3.1, \xi)$ and  $(0, 1+\xi, \xi)$ the scattering is dominated by optic phonon modes. The imaginary part of the dynamical susceptibility obtained from the data of Fig. \ref{Fig3-Sqw_maps} upon subtracting the measured $(\bQ, E)$-independent BG of 4 counts/min and using Eq. \ref{FD_theorem} is shown in the corresponding panels of Fig. \ref{Fig4-Imchi_qw_maps}. It quantifies the system's oscillator response, free of the extra temperature dependence of the intensity resulting from thermal population of excited oscillator states.

It is clear from Fig. \ref{Fig4-Imchi_qw_maps} that no major features appear or disappear upon heating to 300 K and 550 K. The corresponding changes are consistent with modest temperature-induced damping of the phonon oscillator modes, and perhaps with some softening of their energy(ies). However, as shown by the FHWM ellipses in Figures \ref{Fig3-Sqw_maps} and \ref{Fig4-Imchi_qw_maps}, the resolution of the present measurement is too coarse for this issue to be carefully examined.

%
\begin{figure}[!t]
\begin{center}
\includegraphics[width=0.8\linewidth,angle=0]{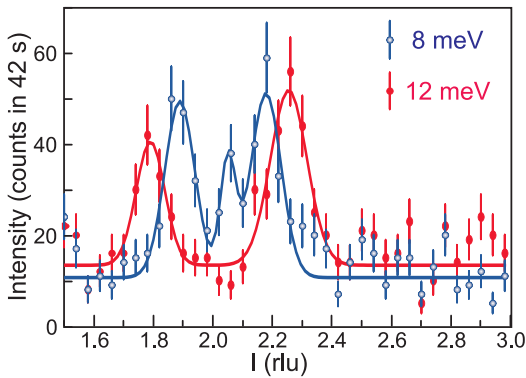}%
\caption{\label{Fig8-002phonon} Constant-energy scans through longitudinal phonon mode near (0,0,2) Bragg reflection at $E = 8$ and 12 meV at T = 6 K, which are part of the contour map shown in Fig. \ref{Fig3-Sqw_maps}, (a). Curves are fits to Gaussian peaks. Small peak at $l \approx 2.05$ in the $E = 8$ meV scan is a tail of the Bragg reflection. }
\end{center}
\end{figure}
%

%
\begin{figure}[!b]
\begin{center}
\includegraphics[width=3.in,angle=0]{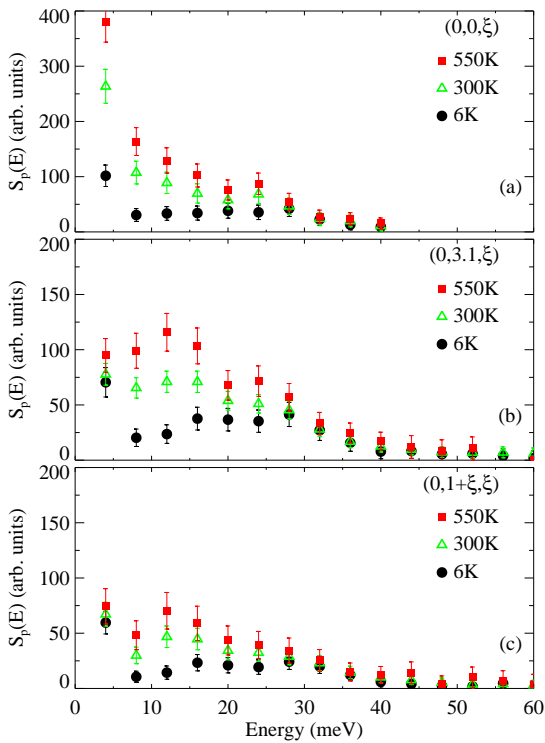}%
\caption{\label{Fig5-Sw} Partial density of states $S_{p}(E)$ obtained by numerical integration of the scattering intensities shown in Figure \ref{Fig3-Sqw_maps}. }
\end{center}
\end{figure}
%

%
\begin{figure}[!b]
\includegraphics[width=3.in,angle=0]{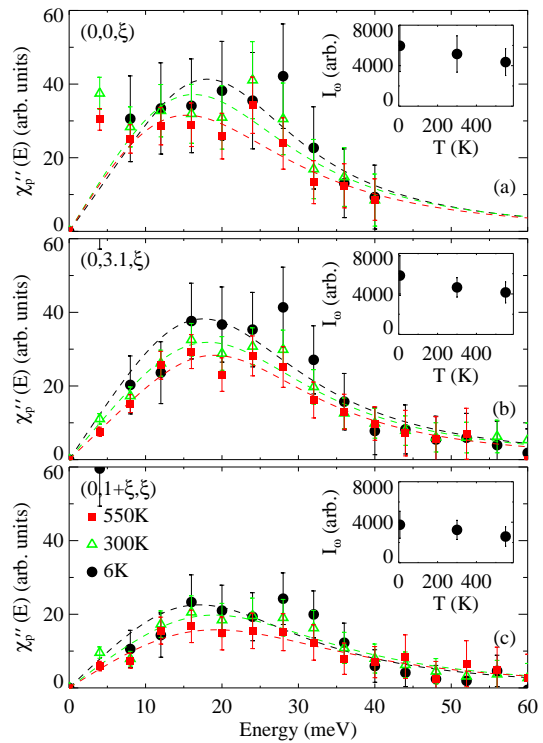}
\caption{\label{Fig6-Imchi_w} Partial $\chi''_{p}(E)$ obtained by numerical integration of $\chi''(\bQ, E)$ shown in Figure \ref{Fig4-Imchi_qw_maps}. Dashed lines are phenomenological fits to $\chi''(E)$ of the damped harmonic oscillator. They primarily serve as guides for the eye, quantifying decrease of $\chi''(\bQ, E)$ with temperature in the range $4\lesssim E \lesssim 60$ meV covered in our measurement. Insets in (a)-(c) show T-dependence of the partial oscillator strength, $I_{\omega} = \int_{5 meV}^{60 meV} E \cdot \chi''(E) dE$ calculated using data in main panels.}
\end{figure}
%

To establish a quantitative detection limit for coherently fluctuating local spin $S_{eff}$, we fit constant-$E$ phonon scans through the longitudinal phonon near the (0,0,2) Bragg reflection to Gaussian profiles, Fig. \ref{Fig8-002phonon}. This yields phonon velocity $v_{ph} = 57.0(3)$ meV/rlu and the energy-integrated intensity of 280(50) cts$\cdot$meV, which corresponds to the calculated cross-section of 110(5) mbarn (see Appendix \ref{cross-section-app} for details). Detection limit of 100 cts$\cdot$meV on the measured energy-integrated intensity of magnetic mode, corresponding to twice the error bar obtained from phonon fits, gives detectable magnetic cross-section of $\geq 39(2)$ mbarn. For $|F_m({\bf q})|^2 \gtrsim 0.4$, which holds for substantial part of our measured data, this leads to a detectable limit of  $S_{eff}^2 \gtrsim 0.25$. Thus, our results establish an upper limit on single-mode magnetic scattering of less than 13\% of the expected magnetic intensity for a singlet-triplet transition in the local-spin picture.

In order to quantify the temperature-induced changes more precisely, we have carried out numerical integration of our data at each energy with respect to the wave vector covered in each map, so as to obtain the partial scattering density of states $S_p(E)$ shown in Fig. \ref{Fig5-Sw}, and the corresponding local dynamical imaginary susceptibility $\chi_p''(E)$ shown in Fig. \ref{Fig6-Imchi_w}. In both cases we observe no indication of coherent peak corresponding to a well-defined localized-spin magnetic excitation. The error bars of our measurement constrain its possible intensity to $\lesssim 20\%$ of the corresponding phonon intensity. It should be mentioned here that in related itinerant-electron pnictides where localized $3d$ magnetic moments are present, such as MnSb and CrSb, intense magnetic excitations on par with phonons were observed by neutron scattering. \cite{Yamaguchi_JMMM1992,RadhakrishnaCable_PRB1996}

Our data, however, do not rule out a possibility that a broad continuum of magnetic excitations of significant integral intensity exists, corresponding to delocalized correlated magnetic states of itinerant electrons. In fact, a hint of such a continuum is contained in the temperature evolution of $\chi_p''(E)$ in Fig. \ref{Fig6-Imchi_w}, which does seem to decrease slightly upon heating in the broad energy range between  5 meV and 35 meV. Although this decrease is not clearly marked outside the statistical error of each given point of our measurement, there seems to be a consistent trend between different points, as well as for $\chi_p''(E)$ corresponding to maps at different \bQ\ shown in panels (a) - (c). It is also supported by phenomenological fits of our data to $\chi''$ of the damped harmonic oscillator (DHO), shown by broken lines interpolating the data points in Fig. \ref{Fig6-Imchi_w}. These fits are purely phenomenological and are used simply to quantify the decrease of $\chi''(\bQ, E)$ with temperature in the range $4\lesssim E \lesssim 60$ meV covered in our measurement. Finally, it could be further quantified by the decrease of the partial (within our energy range) oscillator strength, $I_{\omega}$, which is the quantity involved in the first moment sum rule for the dynamical spin susceptibility. \cite{ZaliznyakLee_2005} $I_{\omega}$ calculated by numerical integration of the corresponding data is shown in the insets in all three panels of the figure. The upper limit on the fluctuating itinerant spin could be estimated from the decrease in the integral intensity of magnetic scattering of roughly $\lesssim 100$ counts$\cdot$meV, which translates into $S_{eff} \lesssim 0.7$, assuming the average magnetic form factor squared of $\approx 0.2$, Fig. \ref{Fig7-FF2}. This is in agreement with the temperature-induced magnetic moment of 1.2$\mu_B$ obtained from static susceptibility measurement. \cite{Petrovic_PRB2003}

If indeed there is  a weak continuum of magnetic scattering in the energy range between $\approx 5$ meV and $\approx 35$ meV, its unambiguous identification is outside the limits of the presently available neutron technology. However, it will become possible in the near future when new high-throughput polarized neutron inelastic spectrometers, such as HYSPEC at the Spallation Neutron Source at Oak ridge in the US, enter operation.

\section{Summary and conclusions}

In summary, we have conducted extensive search for magnetic scattering in the narrow-gap semiconductor FeSb$_2$, which exhibits temperature-induced paramagnetism concomitant with the anisotropic increase of electrical conductivity. The temperature-induced paramagnetic susceptibility and the associated Shottky-like electronic specific heat could be analyzed either in the framework of low (S=0) to high (S=1) spin state transition in the localized spin model, \cite{Petrovic_PRB2003} or as a Pauli susceptibility of itinerant electrons in the itinerant narrow-gap picture similar to FeSi. \cite{Petrovic_PRB2005,Jaccarino_PhysRev1967} A well-defined singlet-triplet magnetic excitation between the crystal field split spin states is expected around $\approx 50$ meV in the localized spin picture. Only a broad continuum of magnetic excitations corresponding to correlated magnetic states of excited itinerant electrons would be expected in the band picture. In both cases magnetic intensity is expected to exhibit characteristic temperature dependence, decreasing upon heating, as excited states get thermally populated.

In our data we find no evidence for a well-defined magnetic excitation corresponding to transitions between the non-magnetic ground state and states of magnetic multiplet in the localized spin picture. We find that peaks in the inelastic response of FeSb$_2$ for energies up to 60 meV and at temperatures $\approx 4.2$ K, $\approx 300$ K and $\approx 550$ K are essentially consistent with the scattering by lattice phonon excitations. Our data establish a quantitative limit of $S_{eff}^2 \lesssim 0.25$ on the fluctuating local spin.

However, a broad magnetic scattering continuum in the 15 meV to 35 meV energy range is not ruled out by our data. In fact, a hint of such a continuum could be traced in the temperature dependence of the imaginary part of the local dynamical susceptibility shown in Fig. \ref{Fig6-Imchi_w}, which decreases upon heating. There should also be an accompanying weak quasielastic paramagnetic scattering similar to that observed in FeSi. \cite{Tajima_PRB1988} The putative magnetic intensity, however, is very weak and its unambiguous identification by the means of polarized neutron scattering would have to be postponed until future developments of the neutron scattering technology.

Our findings make description of FeSb$_2$ in terms of the localized Fe magnetic states unlikely and suggest that paramagnetic susceptibility of itinerant electrons is at the origin of the temperature-induced magnetism in FeSb$_2$.

\begin{acknowledgments}

This work was performed under the Contract DE-\-AC02-\-98CH10886, Division of Material Sciences, US Department of Energy. ATS was funded by National Science Foundation through DMR-0603126. The Research at Oak Ridge National Laboratory's High Flux Isotope Reactor was sponsored by the Scientific User Facilities Division, Office of Basic Energy Sciences, U. S. Department of Energy. ORNL is operated by UT-Battelle, LLC for the U.S. DOE under Contract No. DE-AC05-00OR22725. We thank
M. Lumsden, M. Stone, J.~Tranquada and A. Frenkel for discussions and B. Winn for assistance at HFIR.

\end{acknowledgments}

\begin{appendix}

\section{Magnetic and phonon scattering cross-section}
\label{cross-section-app}%
%
\begin{figure}[!t]
\begin{center}
\includegraphics[width=1.\linewidth,height=1.75in,angle=0]{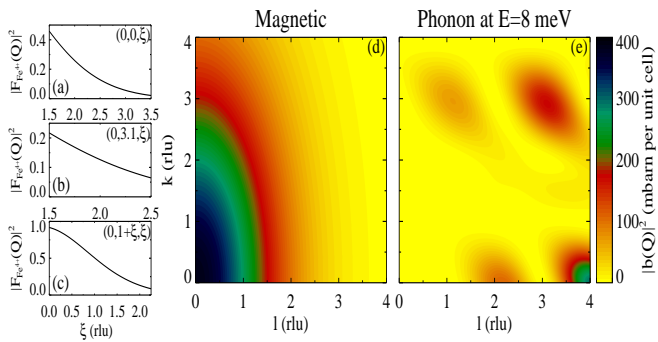}%
\caption{\label{Fig7-FF2} (a) - (c) Fe$^{4+}$ ionic magnetic from factor squared for maps shown in Figures \ref{Fig3-Sqw_maps}, \ref{Fig4-Imchi_qw_maps}. (d) Calculated magnetic scattering cross-section for a non-dispersive, isotropic local triplet mode with $S_{eff} = g/2 \sqrt{S(S+1)} = 1$. (e) Calculated scattering cross-section for a phonon in FeSb$_2$ polarized along the wave vector transfer ${\bf q}$ at $E = 8$ meV, to be compared with panel (d).}
\end{center}
\end{figure}
%

In order to provide quantitative estimates, we compare magnetic scattering expected for a localized, non-dispersive triplet mode corresponding to a singlet-triplet transition at an iron site, to scattering cross-section for the longitudinal acoustic phonon near (0,0,2) Bragg reflection in Fig. \ref{Fig3-Sqw_maps} (a). Magnetic scattering cross-section for a resonant non-dispersive magnetic mode is,
\begin{equation}
\label{sigma_m}
\frac{d^2 \sigma}{d E d\Omega} = 2N \frac{k_f}{k_i} |b_m({\bf q})|^2 \delta(E - \Delta),
\end{equation}
where $k_i$ and $k_f$ are the incident and the scattered neutron wave vectors, $\Delta$ is the mode energy, $N$ is the number of unit cells in the sample (factor 2 accounts for two Fe ions per unit cell), and magnetic scattering length squared is given by,
\begin{equation}
\label{b_m}
|b_m({\bf q})|^2 = r_m^2 |F_m({\bf q})|^2 \frac{2}{3} S_{eff}^2.
\end{equation}
Here $F_m({\bf q})$ is magnetic form factor for the corresponding magnetic ion (we use $|F_m({\bf q})|^2$  for Fe$^{4+}$ for our estimates, Fig. \ref{Fig7-FF2} (a)-(c)), $r_m = -5.39 \cdot 10^{-13}$ cm, and we have introduced the effective spin through $S_{eff} = g/2 \sqrt{S(S+1)}$, where $g$ is the spectroscopic Lande factor. For a singlet-triplet transition, $S_{eff} = g/ \sqrt{2} \approx \sqrt{2}$, and magnetic scattering cross-section is twice that shown in Figure \ref{Fig7-FF2} (d), which was calculated for for $S_{eff} = 1$.

For the long-wavelength acoustic phonon, the scattering cross-section at T = 0 is given by,
\begin{equation}
\label{sigma_ph}
\frac{d^2 \sigma}{d E d\Omega} = N \frac{k_f}{k_i} |b_{ph}({\bf q})|^2 \delta(E - \varepsilon({\bf q})),
\end{equation}
where $\varepsilon({\bf q})$ is its energy, and the scattering length squared is,
\begin{equation}
\label{b_ph}
|b_{ph}({\bf q})|^2 = \frac{\hbar^2 q^2\cos^2\beta}{2 M_{cell} \varepsilon({\bf q})} |F({\bf q})|^2 \approx 2.09 \frac{q^2}{M_{cell} \varepsilon({\bf q})} |F({\bf q})|^2.
\end{equation}
In the latter, $q$ is in $\AA^{-1}$, $M_{cell}$ in atomic mass units, $\varepsilon({\bf q})$ in meV, and $\beta$ is an angle between the phonon polarization vector and the wave vector transfer, ${\bf q}$. $F({\bf q})$ is the unit cell structure factor,
\begin{equation}
\label{F_ph}
F({\bf q}) = \sum_{\mu} b_\mu e^{-i ({\bf q} \cdot {\bf r}_\mu)},
\end{equation}
where $\mu$ indexes atom at a position ${\bf r}_\mu$ in the unit cell, $b_\mu$ is its nuclear scattering length. Scattering cross-section for the longitudinally ($\parallel {\bf q}$) polarized acoustic phonon at $E = 8$ meV, calculated from Eqs. (\ref{sigma_ph}) - (\ref{F_ph}), is shown in Figure \ref{Fig7-FF2} (e). It is already clear from comparison of Fig. \ref{Fig7-FF2} (d) and (e) that in the range where our measurements were performed the expected intensity of the local magnetic mode in the local-spin picture is similar to, or higher than that of the longitudinal phonon in Fig. \ref{Fig3-Sqw_maps}.

\end{appendix}


\end{document}